\documentclass[manuscript]{acmart}
\AtBeginDocument{%
  \providecommand\BibTeX{{%
    \normalfont B\kern-0.5em{\scshape i\kern-0.25em b}\kern-0.8em\TeX}}}

\copyrightyear{2023}
\acmYear{2023}
\setcopyright{acmlicensed}\acmConference[CHI '23]{Proceedings of the 2023 CHI Conference on Human Factors in Computing Systems}{April 23--28, 2023}{Hamburg, Germany}
\acmBooktitle{Proceedings of the 2023 CHI Conference on Human Factors in Computing Systems (CHI '23), April 23--28, 2023, Hamburg, Germany}
\acmPrice{15.00}
\acmDOI{10.1145/3544548.3581053}
\acmISBN{978-1-4503-9421-5/23/04}

\usepackage{url}
\usepackage{enumerate}

\def\markup{0}  % change between 0 and 1 to get the marked-up version and the clean version 
\if\markup 1
\usepackage{soul}
\usepackage[titletoc]{appendix}
\newcommand{\rv}[1]{{\leavevmode\color{blue}#1}}
\else
\newcommand{\rv}[1]{#1}
\newcommand{\st}[1]{}
\fi

\usepackage{graphicx}
\usepackage{xcolor,stfloats}
\usepackage{graphicx}
\begin{document}

% \title{Hindered by Interface, Locked by Algorithm:\newline
% Understanding Blind or Low Vision Streamers' Experience with Douyin Ecosystem\newline
% }

% some other potential titles:

% ``I really, really want my content could reach your feeds":\newline

% ``I really, really want my content could break out of the algorithmic cage":\newline

% ``Please, please don't click `not interested' button if you see my content in your feeds":\newline

%\title[Livestreaming Practices and Challenges Among Streamers Who Are Deaf or Hard of Hearing]{Livestreaming Practices and Challenges Among Streamers Who Are Deaf or Hard of Hearing}

\title[Sparkling Silence: Practices and Challenges of Livestreaming Among Deaf or Hard of Hearing Streamers]{Sparkling Silence: Practices and Challenges of Livestreaming Among Deaf or Hard of Hearing Streamers}

\author{Beiyan Cao}

\affiliation{
 \institution{IIP(Computational Media and Arts)}
  \institution{The Hong Kong University of Science and Technology}
  \city{Hong Kong SAR}
  \country{China}
}
\email{beiyan.cao@connect.ust.hk}

\author{Changyang He}
\affiliation{
  \institution{Department of Computer Science and Engineering, Hong Kong University of Science and Technology}
  \city{Hong Kong SAR}
  \country{China}
}
\email{cheai@cse.ust.hk}

\author{Muzhi Zhou}
\authornote{Corresponding authors}
\affiliation{%
  \institution{Urban Governance and Design,}
  \institution{The Hong Kong University of Science and Technology (Guangzhou)}
  \city{Guangzhou}
  \country{China}
}
\affiliation{
  \institution{Division of Social Science,}
  \institution{The Hong Kong University of Science and Technology}
  \city{Hong Kong SAR}
  \country{China}
}
\email{mzzhou@ust.hk}

% \author{Mingming Fan}
% \affiliation{
%   \institution{Computational Media and Arts Thrust,}
%   \institution{The Hong Kong University of Science and Technology (Guangzhou)}
%   \city{Guangzhou}
%   \country{China}
% }
% \affiliation{
%   \institution{Division of Integrative Systems and Design,}
%   \institution{Department of Computer Science and Engineering,}
%   \institution{The Hong Kong University of Science and Technology}
%   \city{Hong Kong SAR}
%   \country{China}
% }
% \email{mingmingfan@ust.hk}

\author{Mingming Fan}
\authornotemark[1]
\orcid{0000-0002-0356-4712}
\affiliation{
  \institution{Computational Media and Arts Thrust}
  \institution{The Hong Kong University of Science and Technology (Guangzhou)}
  \city{Guangzhou}
  \country{China}
}
\affiliation{
  \institution{Division of Integrative Systems and Design \& Department of Computer Science and Engineering}
  \institution{The Hong Kong University of Science and Technology}
  \city{Hong Kong SAR}
  \country{China}
}
\email{mingmingfan@ust.hk}

\renewcommand{\shortauthors}{CAO, HE, ZHOU, and FAN}

\begin{abstract}
%173 words
%version 1
%Deaf or hard of hearing (DHH) streamers have become increasingly popularity on livestreaming platforms, with some having millions of subscribers. Recent livestreaming studies focused at some minority groups (e.g., people with vision impairments, the LGBT+ community, and disabled video gamers). However, there is limited research on the livestreaming experiences and related challenges among DHH streamers. We conducted semi-structured interviews with 15 DHH streamers to learn why they livestream, the challenges they face and how they navigate software platforms that are often designed for persons who can hear. Our findings revealed the motivations of DHH streamers (e.g., making money, developing social connections, self-presentation and performances, and education.) and strategies they used to mitigate livestreaming difficulties. Sign language, written language, and lip language were utilized to reach accessibility. Diverse kinds of sign language performances were employed to attract more viewers and improve the ornamental value of the performance. As a minority group, DHH streamers encountered technological, social, interactional, and moderation challenges. We provided several suggestions for designing livestreaming platforms to benefit DHH streamers.%

%version 2
Understanding livestream platforms' accessibility challenges for minority groups, such as people with disabilities, is critical to increasing the diversity and inclusion of those platforms. While prior work investigated the experiences of streamers with vision or motor loss, little is known about the experiences of deaf or hard of hearing (DHH) streamers who must work with livestreaming platforms that heavily depend on audio. We conducted semi-structured interviews with DHH streamers to learn why they livestream, how they navigate livestream platforms and related challenges. Our findings revealed their desire to break the stereotypes towards the DHH groups via livestream and the intense interplay between interaction methods, such as sign language, texts, lip language, background music, and viewer characteristics. Major accessibility challenges include the lack of real-time captioning, the small sign language reading window, and misinterpretation of sign language. We present design considerations for improving the accessibility of the livestream platforms.

%Deaf or hard of hearing (DHH) streamers were catching more attention in livestreaming platforms as some of them were having millions of subscribers. Recent livestreaming research studied some minority groups (e.g., blind or low vision, the LGBT+ community, disabled video gamers). However,little research has been done to understand whether and how DHH streamers use livestreaming platforms.We conducted semi-structured interviews with 15 DHH streamers to know what they livestream about, the challenges they face and how they navigate software platforms that are often designed for people who can hear.Our findings uncovered DHH streamers' motivation(e.g., making money, developing social connections, self-presentation and performances, and education.)Accessibility was achieved through the use of sign language, written language, and lip language. To draw in more viewers and increase the ornamental value of the performance, strategies such as diverse types of sign language performances were used. DHH streamers faced technological, sociological, interactional, and moderation difficulties as a minority group. For DHH streamers' convenience,we proposed some design suggestions for livestreaming platforms.

\end{abstract}

\begin{CCSXML}
<ccs2012>
<concept>
<concept_id>10003120.10003121</concept_id>
<concept_desc>Human-centered computing~Human computer interaction (HCI)</concept_desc>
<concept_significance>500</concept_significance>
</concept>

<concept>
<concept_id>10003120.10003121.10011748</concept_id>
<concept_desc>Human-centered computing~Empirical studies in HCI</concept_desc>
<concept_significance>300</concept_significance>
</concept>
</ccs2012>
\end{CCSXML}

\ccsdesc[500]{Human-centered computing~Human computer interaction (HCI)}
\ccsdesc[300]{Human-centered computing~Empirical studies in HCI}

\keywords{deaf or hard of hearing (DHH), livestreaming, accessibility, social media/online communities}

\maketitle

\section{INTRODUCTION}
\label{INTRODUCTION}

Online livestreaming is a booming industry in countries like the US and China. In recent years, a growing number of online platforms, such as Tiktok, Twitch, and Youtube offer livestreaming services. More people are attracted to livestream on various platforms. In China, the number of users reached 617 million in 2020 \cite{r2}. Among them, many are deaf or hard of hearing (DHH) streamers. Hearing loss or difficulties in hearing is not only one of the disabilities that many people carry with them for most of their lives, but we all have the chance to be DHH as we grow older. By the year 2050, approximately 2.5 billion people in the world will suffer from varying degrees of hearing loss \cite{wrd2021}. In China, more than 14 million males have hearing loss, while 13 million women are DHH \cite{r1}. Their demand for livestream and the impact they have made on livestreaming platforms should not be ignored. Some leading DHH livestreamers have already had millions of subscribers \cite{DHHOnline}. 

Although most studies are about the livestreaming experiences of able-bodied people, a limited number of studies have reported the livestreaming experiences of people who are blind or have low vision \cite{Rong2022, Jun2021}. In one study about the setting of video calls for work where DHH people are included, the importance of the image quality is highlighted \cite{Tang2021}. So far as we know, little research has investigated DHH people’s livestreaming experiences about how they navigate the platform and interact with other streamers and viewers. People who are deaf or hard of hearing may have their unique ways of livestreaming practices due to their own form of communication, culture, and identity. In particular, the communication barrier between the DHH group who uses sign language and the majority of the livestream platform users may bring new challenges for livestreaming platforms that rely heavily on audio input. This situation is even more pronounced in China because Chinese DHH groups mainly rely on natural sign languages for daily communication. Natural sign language in China grows from people's daily activities and is embedded in the local culture. Accordingly, the natural sign languages in different areas are different, similar to the many dialects in China. Therefore, communication even within the DHH group is not without barriers. We should seek new opportunities for improving their livestreaming experience and identify ways of improving the accessibility of livestreaming platforms. This paper will fill this gap by investigating DHH streamers' livestreaming experiences, including their motivations and content of livestreaming, how they interact with viewers and other streamers, and what challenges they have. 

We aim to identify accessibility issues of the livestream platforms for DHH streamers and offer useful suggestions to overcome these difficulties and challenges. To achieve this goal, we conducted semi-structured interviews with 15 DHH livestreamers in China, who used a diverse range of livestreaming platforms. We summarize our research questions below:

\begin{itemize}
    \item RQ1: What motivates DHH people to livestream?
    \item RQ2: What are the practices of DHH streamers? Specifically, how do DHH streamers interact with their viewers and other streamers on livestreaming platforms?
    \item RQ3: What challenges do DHH streamers face? 
\end{itemize}

Our interviews revealed that besides the commonly known reasons for livestreaming, such as monetary gain or social demands, many streamers believed that their livestream can break stereotypes towards the DHH group. DHH streamers also employed a wide range of interaction methods, such as sign language, texts, 
lip language (lip reading and speaking), background music, or the combination of them to meet different interaction needs. One important finding was that the selection of which method to interact was dependent on and was reinforced by whether the viewers were DHH or not. This intensive interaction between communication method and viewer characteristics had important implications to the livestreaming content and experiences. They also employed different strategies to mitigate various accessibility issues. In the end, they shared with us their hopes and expectations that the livestreaming platforms can improve.    

As one of the first studies to examine how DHH streamers interact with viewers and other streamers (co-performers who livestream together) and how they navigate the livestream platforms, our work contributes to the current literature on accessibility by uncovering the lived experiences of a group of people rarely considered in the design and development of online livestreaming platforms. Their unique ways of communication, their sense of identity, stereotypes towards this group, as well as the technical shortcomings that heavily rely on audio all shape the interaction patterns of DHH streamers on livestream platforms. At the end of this paper, we make some suggestions for designing a more inclusive and equitable livestreaming platform environment to enhance the user experience of DHH streamers.

%a few studies have reported how people who are blind or have low vision livestream \cite{Rong2022,Jun2021}. We still know very little about the livestreaming experiences of those who are deaf or hard of hearing (DHH). People of different disabilities face different challenges when they livestreaming. We should seek new opportunities for improving their livestreaming experience and identify ways of improving the accessibility of livestreaming platform so that more people can enjoy and benefit from livestream services. In this paper, we focus on the group of streamers who are deaf or hard of hearing.% 

\section{RELATED WORK}
\label{RELATED WORK}

\subsection{Livestreaming and Its Practice in China}

% Prior work has investigated the broad use of livestreaming across different topics. One line of work looked at the e Another strand of work focused on how l  Also, when lacking the nature of keeping audience engaged compared to entertainment-centered livestreaming~\cite{fraser2019sharing,yang2020snapstream} These work also demonstrated how streamers applied interactional strategies to keep and promote viewers' engagement.  and cultivate a collective entertainment atmosphere

Livestreaming, as a medium that enables synchronous communication of practice, knowledge and opinions, has attracted increasing research attention. Entertainment-centered livestreaming, such as those that stream games~\cite{pellicone2017game,hamari2017esports,hamilton2014streaming}, live events~\cite{tang2017crowdcasting}, and outdoor activities~\cite{lu2019vicariously}, has been shown to foster participatory entertainment experiences in the livestreaming community~\cite{hamilton2014streaming, li2019live}. Livestreaming also enables knowledge sharing. Previous work has shown how livestreaming programming skills~\cite{haaranen2017programming,faas2018watch}, language learning~\cite{samat2019live}, digital art making~\cite{fraser2019sharing}, and cultural practices~\cite{lu2019feel} have great potential to facilitate online education and learning. These studies also highlight how streamers use different interaction strategies, such as being responsive in the stream chat~\cite{faas2018watch} and inviting audiences for performance~\cite{li2019live}, to maintain and encourage viewer engagement. 

% Under the popularity of livestreaming across the world, a; market size as of; HCI and CSCW researchers have explored u

The livestreaming industry is also booming in China, with more than 600 million users and a market value of \$28.7 billion in 2020~\cite{li2022China}. In China, the two most famous livestreaming apps are Douyin  (the Chinese version of Tiktok) and Kuaishou (it means “fast hands”). The distinct livestreaming content and practices in China against the backdrop of the social and cultural context have attracted scholars' attention~\cite{lu2018you,lu2019feel,lu2019vicariously,lu2021more,tang2022dare,li2019live}. For instance, Lu et al. reported how Intangible Cultural Heritage (ICH) activities are showcased in livestreaming~\cite{lu2019feel}. Virtual streamers, in the form of 2D or 3D avatars with human voices, are also popular in China~\cite{lu2021more}. 

Co-performance was popular on livestreaming platforms~\cite{li2019live}. In co-performance, streamers livestreamed together, and all their viewers could watch all connected streamers. Streamers and audiences often collaborated to deliver livestreaming performances. \textit{Lianmai} and \textit{PK} are the two common forms of co-performance. \textit{Lianmai} is similar to an online video call, where multiple streamers are connected. \textit{PK} is more entertaining than \textit{lianmai}. In \textit{PK}, two streamers are connected to compete for a small task and their viewers can vote which one perform better in a given time. The interface is shown in Fig. \ref{fig:lianmaiVSPK}, with a video-connection interface (i.e., \textit{lianmai}), spectators can present their live performance through audience windows embedded in streamers' streaming video~\cite{li2019live}.

Recent work also looks at the opportunities and challenges that livestreaming brings to socially marginalized groups in China. For example, rural Chinese female streamers gained a sense of self-empowerment through livestreaming-based e-commerce but suffered social sanctions due to their deviation from the traditional gender roles embedded in the Chinese society~\cite{tang2022dare}.

The widespread adoption of livestreaming in China also provides a unique opportunity for another marginalized group---the 27-million DHH people~\cite{r2}---to express themselves and connect with others. However, the current livestreaming platform is designed for speaking language communication. Sign language communication, one of the major ways of communication for DHH people, places new demands and challenges on the existing livestreaming platform. However, we know very little about what technical challenges the DHH livestreaming community faces. Besides, as a distinct and underrepresented group in the livestreaming community, it is also worth exploring whether and how identity-related issues affect DHH streamers' practices and perceptions. Our work will fill a significant research gap in HCI by examining what challenges and opportunities livestreaming presents for DHH streamers.

% Marginalized
% With a of 20 million .

% This work fills a significant research gap in HCI by investigating how livestreaming empowers DHH populations, and how this group .

% When livestreaming is featured in . 

\subsection{Inclusive and Accessible Livestreaming}

% The  high internet penetration in China (xx\% in 2021~/cite{???}) and the relatively low prices of digital devices such as smartphones make watching and doing livestreaming convenient. Therefore,

Livestreaming is an increasingly popular channel that socially marginalized populations such as older adults~\cite{struzek2019live,chang2021effectiveness}, low-income groups~\cite{peng2021optimal, tang2022dare}, and people who are Blind or of Low Vision (BLV)~\cite{jun2021exploring, Rong2022} use to present themselves, communicate with others, and earn money. However, these marginalized groups have long suffered from various social stigmas that categorize them as weak, vulnerable, or pathetic, and they may face negative stereotypes as streamers, such as receiving ironic comments~\cite{tang2022dare}. Some streamers in marginalized groups also interact in a way that stands out from mainstream streaming practices, such as using sign languages for communication among DHH streamers. A more inclusive design is therefore needed to support a wider range of communication practices and make livestreaming more accessible.

% For example, Lu et al. noted that older adults were attracted to watch knowledge and experience sharing streams, and urged for more inclusive design to support their social interactions~\cite{lu2018you}. Peng et al. found that livestreaming has been a significant approach for low-income farmers to promote farm products sales~\cite{peng2021optimal}. Nevertheless, social stigmas appear to be a critical challenge for marginalized groups in the livestreaming community~\cite{tang2022dare}. How livestreaming could be more inclusive and accessible is a crucial research topic.

% though the nature of multi-modality synchronous communication affords accessibility ,

% In addition to the negative stereotypes that socially marginalized groups often face online, DHH or BLV groups have other challenges when navigating livestream platforms, 

% especially when the consideration of specific communications and interactions of these groups is still on the fringe of current livestreaming design. For example, compared to non-DHH users, users using sign language may expect higher video quality to facilitate synchronous streamer-spectator communication. 

% Unfortunately, the consideration of specific communications and interactions of these groups is still on the fringe of current livestreaming design. 

Recent studies have started to focus on the specific experiences and challenges of the socially marginalized populations in livestreaming (e.g., BLV streamers~\cite{jun2021exploring,Rong2022}), unpacking their nuanced communication and interaction needs that provide insight into accessible livestreaming design. For example, Jun et al. found that streamers with visual loss not only have barriers to multitasking while streaming, but also experience difficulties keeping up with the live chat~\cite{jun2021exploring}. Rong et al. further identified that BLV streamers tend to be trapped in the BLV livestreaming community when their livestreaming content is disproportionally recommended to BLV viewers due to algorithmic biases~\cite{Rong2022}.

To date, little work has investigated the livestreaming practices of the group of DHH streamers. Our work presents one of the first attempts to investigate the distinct livestreaming experiences of this group and shed light on new ways to make livestreaming more accessible.

% sign language

% With a relatively low threshold to use, livestreaming has been an channel that underrepresented and marginalized populations accessible to  interactional for . For instance, .

\subsection{Accessibility of Information and Communication Technologies for DHH users}

A large body of work has underscored the importance of providing accessible Information and Communication Technologies (ICTs) to facilitate the adoption and use of ICTs for the DHH population (e.g., ~\cite{mack2020social,lee2021american,rui2022online,seita2022remotely,glasser2022analyzing,li2022exploration}). Kožuh and Debevc showed that the perceived accessibility of ICT is strongly correlated with its use in the DHH population~\cite{kovzuh2018challenges}. Several techniques to enhance accessibility at the application design/development level have been proposed and evaluated. These include providing non-audible feedback~\cite{schefer2018supporting}, sign language support~\cite{alnfiai2017social}, and speech-to-text translation~\cite{chiu2010essential}. Machine-generated or user-generated content, especially captions, has been shown to further improve accessibility for DHH internet users~\cite{shiver2015evaluating,kawas2016improving,li2022exploration,mcdonnell2021social}. We know that both caption accuracy~\cite{kawas2016improving,li2022exploration} and caption rate~\cite{tyler2009effect} have a strong impact on content comprehension. Unfortunately, many DHH users struggle with low-quality captions on video-sharing platforms~\cite{li2022exploration}. Meanwhile, as warned by Alnifial et al., most apps largely missed several important features desired by DHH users, such as supporting sign language and affording text-to-speech interfaces~\cite{alnfiai2017social}. 

%  For example, Shiver and Wolfe found that DHH individuals generally expressed the significance and usefulness of having captions when watching online videos~\cite{shiver2015evaluating}. Existing literature also calls for a more holistic view for DHH populations, which could be crucial in understanding DHH groups' perceptions on ICT use and designing accessible technologies. The  construction of disability potentially brought bad solutions to real problems due to its misunderstanding nature~\cite{lane2005ethnicity}. 

% In particular, The reason is that m

% The DHH group also differs from groups with other forms of disabilities. 

Scholars have also called for a more holistic view of the needs of the DHH population, especially considering their social and cultural identities, in designing accessible ICTs for them. As argued by Harlan Lane, deafness has become an identity similar to ethnicity rather than just one type of disability~\cite{lane2005ethnicity}. Language minorities using sign languages have their own linguistic, conversational, and cultural norms~\cite{lane2005ethnicity}. Similarly, McKee et al. highlighted that the deaf community views deafness as a cultural identity, rather than a disability~\cite{mckee2013ethical}. On this note, Hermawati and Pieri suggested that studying the acceptance and development of assistive technologies should focus on the overall welfare of DHH populations beyond only hearing-related challenges~\cite{hermawati2019assistive}, such as social and cultural norms in DHH online communities.

In this study, we consider those ICT-related developments and challenges related to the DHH group and further focus on the specific setting of livestreaming platforms. We adopt the above identity approach to build our research questions and interpretations of some of the findings. This work contributes to the literature on the accessibility of livestreaming platforms by examining the motivations, practices, and challenges of DHH streamers. Different from the curated online videos where high-quality captions can be prepared, the livestreaming setting is featured by synchronicity and thus requires a reassessment of accessibility for DHH online users. This makes facilitating synchronous and interactive sign language communication a key challenge. Through the identity perspective discussed above, we also relate the social and cultural characteristics of DHH groups to their interaction and perceptions of livestreaming. This study is one of the first to investigate DHH streamers' experiences and provides important insights for developing DHH-friendly livestreaming services.

% , and even machine-generated captions that contain errors can contribute to their understanding

\begin{table*}[ht]

\caption{Summary of DHH livestreamers interviewed. Among the 15 participants, 8 were female and 7 were male, aged from 20 to 40.DHH state\uppercase\expandafter{\romannumeral1}——Speech recognition rate \textless15\%; DHH state \uppercase\expandafter{\romannumeral2}——Speech recognition rate 15\%-30\%; DHH state \uppercase\expandafter{\romannumeral3}——Speech recognition rate 30\%-60\%; DHH state \uppercase\expandafter{\romannumeral4}——Speech recognition rate 61\%-70\%;}

\scalebox{1}{
\begin{tabular}{c|c|c|c|c|c|c}
    \toprule
    ID    & Gender & Age   & Platforms in Use & Livestreaming History & Livestreaming Content & \multicolumn{1}{c}{DHH State} \\
    \midrule
    P1    & M     & 22    & Kuaishou & 1 week & Education(Sign Language) & \uppercase\expandafter{\romannumeral2} \\ \hline
    P2    & M     & 22    & Tencent Meeting; Bilibili & 1 year & Education & \uppercase\expandafter{\romannumeral1} \\ \hline
    P3    & F     & 22    & Kuaishou & 2 months & Education(Sign Language) & \uppercase\expandafter{\romannumeral3} \\ \hline
    P4    & M     & 28    & Kuaishou & 1 month & Dance \& Sign Language Hiphop & \uppercase\expandafter{\romannumeral3} \\ \hline
    P5    & F     & 24    & Wechat Group Live & 2 weeks & Freechat & \uppercase\expandafter{\romannumeral4} \\ \hline
    P6    & F     & 20    & Bilibili & half a year & Freechat \& Sign Language Song & \uppercase\expandafter{\romannumeral3} \\ \hline
    P7    & F     & 29    & Douyin & 1 year & Freechat & \uppercase\expandafter{\romannumeral1} \\ \hline
    P8    & F     & 30    & Kuaishou & 1 year & E- commerce & \uppercase\expandafter{\romannumeral2} \\ \hline
    P9    & F     & 24    & Kuaishou & 4 years & Business/work & \uppercase\expandafter{\romannumeral2} \\ \hline
    P10   & M     & 25    & Huya  & 4 years & Game  & \uppercase\expandafter{\romannumeral2} \\ \hline
    P11   & M     & 24    & Bilibili & half a year & Game \& Sign Language Song & \uppercase\expandafter{\romannumeral1} \\ \hline
    P12   & M     & 24    & Kuaishou & 3 years & Magic Tricks \& E- commerce & \uppercase\expandafter{\romannumeral2} \\ \hline
    P13   & F     & 29    & Douyin & 2 years & Freechat & \uppercase\expandafter{\romannumeral2} \\ \hline
    P14   & F     & 40    & Kuaishou & 3 years & E- commerce & \uppercase\expandafter{\romannumeral1} \\ \hline
    P15   & M     & 35    & Kuaishou & 2 years & E- commerce & \uppercase\expandafter{\romannumeral4} \\ \hline
    \end{tabular}%
}
\small
\centering
\Description[Summary of DHH livestreamers interviewed]{A table with 7 columns and 16 rows. Participants' Information is given, including ID, gender, hearing condition, age, platforms in use,livestreaming history and livestreaming content.There were eight females and seven males. Four~(N=4) participants were in \uppercase\expandafter{\romannumeral1} DHH level (Speech recognition rate \textless15\%) , six~(N=6) participants were in \uppercase\expandafter{\romannumeral2} DHH level (Speech recognition rate 15\%-30\%) , three~(N=3) participants were in \uppercase\expandafter{\romannumeral3} DHH level (Speech recognition rate 30\%-60\%), and two~(N=2) participants were in \uppercase\expandafter{\romannumeral4} DHH level (Speech recognition rate 61\%-70\%) . Participants were between 20 and 40 years old~($M=27, SD=5.42$) with livestreaming history ranged from 1 week to 4 years ($M=77.27, SD=74.08$,unit=weeks). Eight livestreamed on Kuaishou, 3 livestreamed on Bilibili(one of them also used Tencent Meeting), 2 livestreamed on Douyin, 1 livestreamed on Wechat and 1 livestreamed on Huya.  Livestreaming content is also very rich and includes educational content, e-commerce, free chat and some talent shows such as games, magic shows, sign language and songs.}

\label{table:participants}
\end{table*}

\section{METHOD}
\label{METHOD}

We conducted semi-structured interviews with 15 DHH streamers, combined with a brief pre-interview survey. In the survey, we collected basic information from the DHH streamers, such as their age, gender, DHH condition (this state follows the DHH classification of Chinese Practical Assessment Standard for Disabled Persons \cite{DHHstandard}), the main livestreaming platform used and their user names and livestreaming history. Based on this information, we prepared more appropriate questions to assist the interview. 

\subsection{Participants}
\label{Participants}
We recruited participants by distributing posters with our contact information on Xiaohongshu or by snowball sampling. Xiaohongshu is a social media platform that has gained high popularity in China in recent years by providing an avenue for sharing curated short-videos, photos, or images. Six DHH streamers contacted us directly for an interview after reading our posters. We recruited another nine participants through snowball sampling. Our respondents are from a diverse background in respect of their DHH states, the livestreaming platforms they use, livestreaming content, and  livestreaming history.

Based on the information provided in the pre-interview survey, we located and viewed the participants’ profiles on their livestreaming platforms to gather further information such as the number of subscribers and the content of their published short videos. This process helps us to design more relevant questions for semi-structured interviews. 

Table 1 shows a summary of participants' information, including their gender, age, platforms, livestreaming history, livestreaming content and hearing conditions. There were eight females and seven males. Four~(N=4) participants were in DHH level \uppercase\expandafter{\romannumeral1} (Speech recognition rate \textless15\%) , six~(N=6) participants were in DHH level \uppercase\expandafter{\romannumeral2} (Speech recognition rate 15\%-30\%) , three~(N=3) participants were in DHH level \uppercase\expandafter{\romannumeral3} (Speech recognition rate 30\%-60\%), and two~(N=2) participants were in DHH level \uppercase\expandafter{\romannumeral4} (Speech recognition rate 61\%-70\%). Participants were between 20 and 40 years old~($M=27, SD=5.42$) with livestreaming history ranged from 1 week to 4 years ($M=77.27, SD=74.08$, unit=weeks). %14 of them use sign language in daily livestreaming.% 
Eight livestreamed on Kuaishou, 3 livestreamed on Bilibili(one of them also used Tencent Meeting), 2 livestreamed on Douyin, 1 livestreamed on Wechat and 1 livestreamed on Huya. Livestreaming content is also very rich and includes education, e-commerce, free chat and some talent shows such as games, magic shows, sign language and songs.

\subsection{Procedure}
\label{Procedure}
%After the interview, each participant received 80 CNY \rv {(about 14 dollars)}. The online meetings were conducted in Chinese sign language or Putonghua spoken language and video recorded. The text interviews were in Chinese and automatically recorded.

Our goal was to uncover the motivations, practices, and challenges of livestreaming among DHH streamers. We obtained participants' informed consent prior to the interviews.
Our interviews were conducted online between April 19 and June 17, 2022. 
Participants could choose to be interviewed via online text messages, online meetings with communication in sign language with the assistance of an interpreter (this design may impact our findings, please see our later discussions in the limitation session), or any other forms that they preferred. Interestingly, five of the participants opted to livestream the interviews on Kuaishou. Ten participants were interviewed via online text messages on WeChat. All interviews lasted between one and three hours and were video recorded and transcribed for later analysis. The different modes of the interview may affect the interaction between the interviewer and the participants and the findings. When we report the results later in 4. Finding section, we did not observe that our results are significantly affected by the mode of interview. The reason could be that our interviews were about their general livestreaming experiences (see below) and did not involve highly sensitive questions.

During the interview, we first asked why they wanted to livestream and what motivated them to continue doing so. We then asked about the frequency and duration of livestreaming, what they livestream, and how they interact with their viewers. We focused on challenges that they have encountered and whether they have been able to resolve those issues, and what measures they have taken to enhance their livestreaming experiences. We also asked what they hope the livestreaming platforms can do to improve accessibility. 
After the interview, each participant was compensated 80 CNY \rv {(about 14 dollars)}.

\subsection{Data Analysis}
\label{Data Analysis}
Our data included audio recordings of online meetings and text messages from WeChat. The audio recordings were first transcribed into a text script. Four researchers, who are native Chinese speakers, firstly read through the text script several times to have an overall understanding of the livestream experiences of DHH streamers. Then, two of the researchers independently coded the script using an open-coding approach~\cite{corbin2014basics}. Deductive and inductive coding techniques were combined. First, we have established three main themes—motivation, practice, and challenges—to guide the semi-structured interview. Within each main theme, the sub-themes and the specific contents were inductively constructed by assigning keywords to the answers provided by the participants. Repeating or similar keywords are grouped at a higher level. For instance, when participants explained what motivated them to start livestreaming, we would label that part as the main theme "motivation". Then, the sub-theme "Economic Benefits" was identified when the words "money," "income," and "online sale/commerce" appeared in the responses. In this process, the two coders regularly discussed the codes and resolved disagreements to create a consolidated codebook. Further meetings were scheduled with all co-authors to reach agreements based on the preliminary coding result.
Finally, we synthesize our tags surrounding the three themes in the research questions. They are \textit{motivations}, \textit{practices}, and \textit{challenges}.

\begin{figure*}
    \centering
    \includegraphics[width=0.95\textwidth]{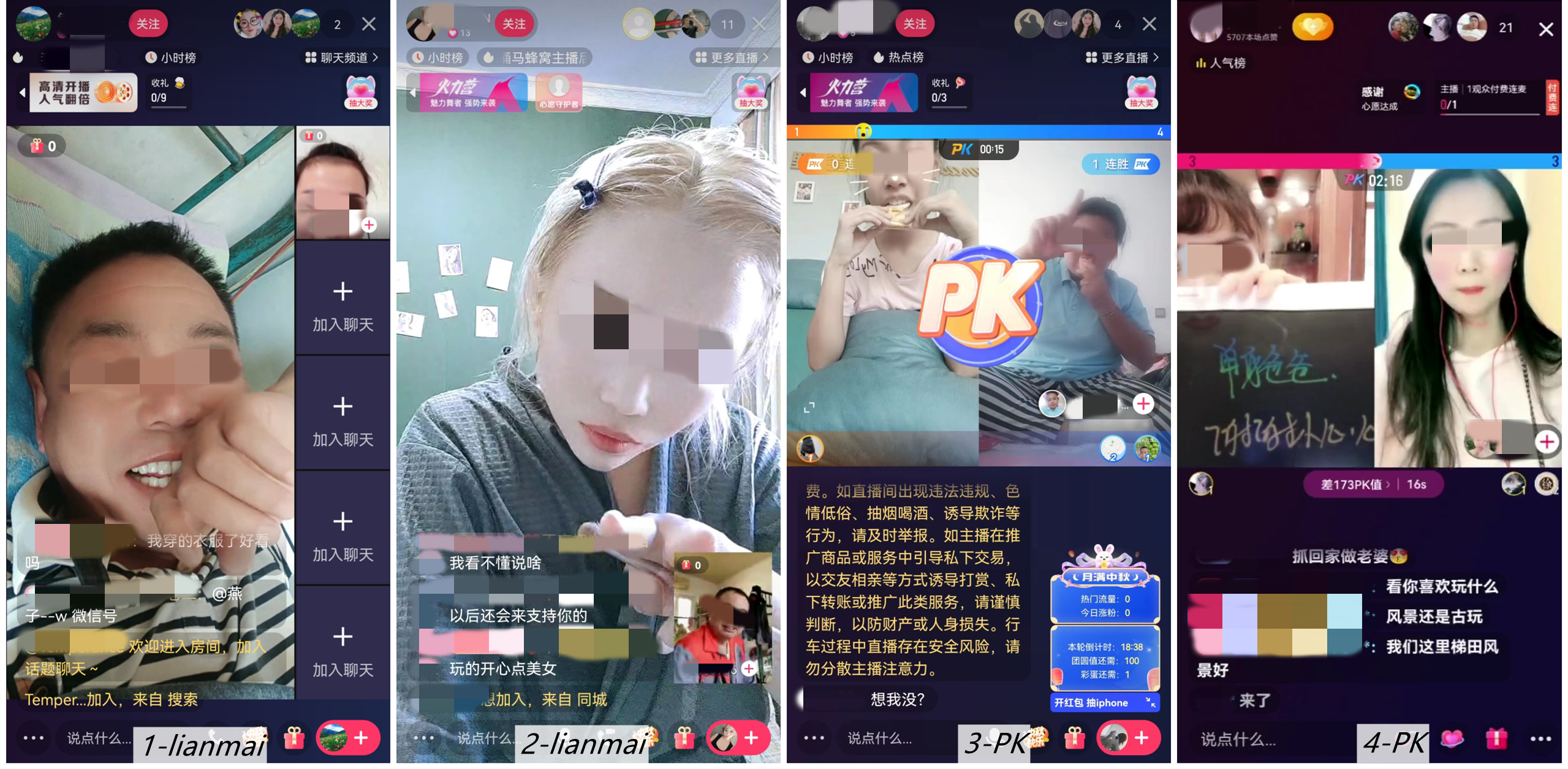}
    \Description[Screenshot of DHH streamers' two co-performance format:\textbf{lianmai} and \textbf{PK}]{four captures separately showed DHH livestreaming co-performance of DHH streamers.DHH streamers co-perform on Douyin and Kuaishou.The left two screenshots show DHH streamers \textbf{lianmai}(one kind of co-performance) in sign language with separated one big window and one or several small ones and no time restrictions while the right two screenshots showed DHH streamers\textbf{PK} (the other kind of co-performance) with other streamers.The right most 4-PK capture show one DHH streamer is using written mandarin on a writing pad with each having half of the window size and remaining time two minutes and 16 seconds}
   \caption{DHH streamers co-perform on Douyin and Kuaishou.The left two pictures show DHH streamers \textit{lianmai} in sign language with separated one big window and one or several small ones for the guest streamers while the right two pictures shows DHH streamers \textit{PK} with other streamers.The right most picture (4-\textit{PK}) captures one DHH streamer who is using a writing pad.}
   \label{fig:lianmaiVSPK}
\end{figure*}

\section{FINDINGS}
\label{Findings}
%(e.g., economical benefits, social demands, self-presentation and performances, as well as DHH ulture and identity promotion.) For example,Sign language, written language, and lip language were utilized to reach accessibility.

In this section, we describe the findings on motivations, practices and challenges of livestreaming among DHH streamers. We first summarize four major reasons that motivate DHH streamers to livestream in Section \ref{findings: RQ1} (RQ1). In Section \ref{findings: RQ2}, we provide a detailed illustration of how DHH streamers navigated the livestreaming platforms, how they interacted with their viewers and other co-performing streamers who might or might not be DHH, and their strategies to engaged viewers (RQ2). We finally present four types of livestreaming challenges, linking to technology, interaction, content moderation, and social stereotypes in Section \ref{findings: RQ3} (RQ3). Please note that most of our participants become DHH from a very young age, communicate primarily through sign language, and are therefore considered culturally deaf.

\subsection{Motivations and Contents (RQ1)}\label{findings: RQ1}

 Because livestreaming motivations were closely linked with what they livestream, we introduced the two together in this section. DHH streamers livestreamed for reasons similar to non-DHH streamers. We found that they wanted to make money from livestreaming e-commerce, to meet their societal demands, or to showcase their talents. But, one frequently mentioned distinct motivation was to change the stereotype about people who were DHH. Economic benefits and positive feedbacks from viewers were the two most frequently cited reasons that drive DHH streamers to continue livestreaming. Of course, many streamers mentioned multiple reasons.

\subsubsection{\textbf{Economical Benefits}}

Economic gains were one of the most commonly referred reasons. Most livestreaming platforms allowed viewers to reward their favorite streamers by purchasing gifts, props, etc.. These gifts and props could be directly exchanged into money to steamers' e-wallets. In addition, streamers in e-commerce could provide external links that allow viewers to purchase goods advertised in the livestream. The streamers could then receive advertising fees or service fees from the third parties selling or producing these goods.

Livestreaming was two DHH streamers’ job duty. P3 worked for a company that made captioning eyeglasses for DHH people. P3 spoke Putonghua (the standard spoken form of Chinese) and also could use sign language for communication. Her employer asked her to livestream to find potential clients, who could be either DHH or non-DHH viewers. P4, who had severe hearing loss, was an employee of a sign language culture Co, LTD. Disseminating sign language culture was one of his job responsibilities. He conveyed hip-hop songs in sign language and had made a name for himself on Kuaishou.

%P12 was a magician who gave magic performances all around China. The COVID-19 pandemic halted his travel plans and ruined his source of income. He switched his magic shows from offline to online. By doing so, his financial situation improved substantially.

Five (P8, P11, P12, P13, P15) of the participants were self-motivated to seek economic benefits from livestreaming. P8 was inspired by other DHH streamers,\textit{``Because I saw that many DHH streamers make a lot of money, so I want to try it too.’’} P13 also said,\textit{``Livestream platforms offer high traffic with a low threshold. This makes it easier for me to start.''} P11 was a part-time streamer who accidentally received a virtual gift from his viewer and found that livestreaming can make extra income,\textit{``so I continued to do it.''}

\subsubsection{\textbf{Social Demands}}

Many DHH streamers mentioned that they livestream because of the social support they have gained on various platforms. P9, a 24-year old girl who worked far away from her family, said,\textit{``When I livestream, my grandpa and my parents can see my face and know how am I doing recently. It makes them very happy.’’} By livestreaming regularly, her family can find out about her current living condition, which enhances the solidarity of her family. Similar to P9, P5 also mentioned that \textit{``My friends said when they saw me livestream, they felt reassuring and happy.''}

In addition to the chance to receive positive feedbacks from friends and families, some livestreamed because they wanted to improve their social skills. P1 said \textit{``I want to practice my social skills and make more friends.''} P8 benefited greatly by socializing via livestream---she turned more outgoing.

%\begin{quote}
    %\emph{``I used to be very introverted and didn't like to deal with people who are not DHH  or people that I am not familiar with. I used to be at home all day and didn't go out. Now I have slowly overcome my fears and learnt to communicate with different people. I have made many friends and often talk with them.''}
%\end{quote}

\subsubsection{\textbf{Self-Presentation and Performance }}
%\label{}
\begin{figure*}
    \centering
    \includegraphics[width=0.95\textwidth]{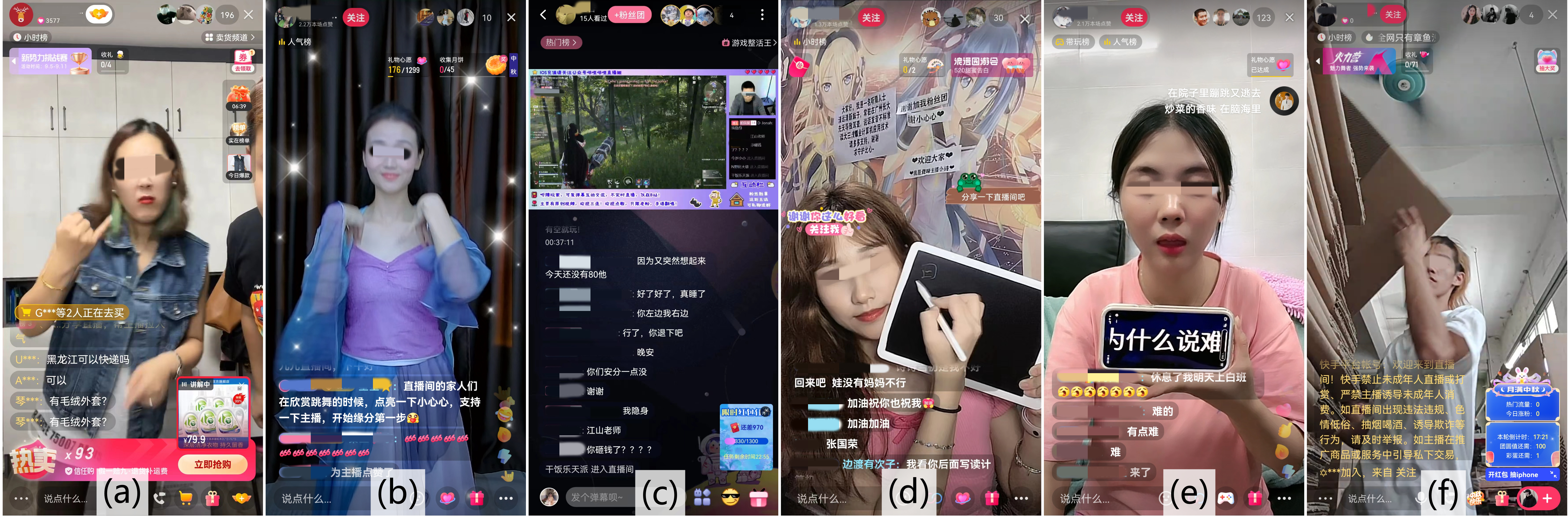}
    \Description[Screenshot of DHH streamers' livestreaming]{Six captures separately showed various DHH livestreaming content.DHH streamers livestream on Douyin, Kuaishou and Bilibili. (a) a DHH e-commerce streamer introducing a product using sign language, (b) dancing, (c) playing video games, (d) Chatting with viewers by writing on a pad, (e) Chatting with viewers by texting on a smartphone and displaying it, (f) working (e.g., handling cartons in a factory).}
   \caption{DHH streamers livestream on Douyin, Kuaishou and Bilibili. (a) a DHH e-commerce streamer introducing a product using sign language, (b) dancing, (c) playing video games, (d) Chatting with viewers by writing on a pad, (e) Chatting with viewers by texting on a smartphone and displaying it, (f) working (e.g., handling cartons in a factory).}
   \label{fig:DHHIntro6}
\end{figure*}

Another important reason was to showcase their talents, skills, and insights into daily life to earn some level of social recognition. P13 loved sharing her life experiences such as traveling with her viewers, and P9 preferred to share her working experiences of running a nail salon. The magician (P12), who switched his magic show from offline to online, said, \textit{``I love magic. I want to share it with my viewers!’’}. P4, who performed hip-hop in sign language, said, \textit{"I love dancing and performing sign language songs. I want more people to know about sign language songs."} Within one month, he gained more than 700 subscribers on Kuaishou. In addition, P6 and P11 collaborated ("\textit{lianmai}",the first two pictures of Fig. \ref{fig:DHHIntro6}) on songs in sign language when livestream. They became friends after meeting in a sign language class. Before they started livestreaming, they posted short videos and gained some subscribers. \textit{``Some subscribers want me to perform a song in sign language through livestream so that they can see whether my real-time performance is as good as in my curated video posts''}, P11 said. P11 also wanted to change people’s views towards sign language songs. \textit{"People always think that sign language songs are slow songs. However, I think sign language songs can also be fast. That’s why I like to play rock and roll."}

Those who engaged in self-presentation and performance seemed to have a deeper reason that drives them to continue livestreaming — changing the stereotype towards DHH people. P10, who was an excellent video game player, when asked about why he started livestreaming playing video game, he replied\textit{``I want the public to know that people with hearing loss can be very good at playing video games!''} P11, who was also a video game streamer, also expressed similar thoughts. P14 was a makeup streamers who always showed pretty in front of the camera, and shared her cosmetics and outfits with her viewers. \textit{``We can be pretty too''}, she said. In summary, by showing their talents, skills, and the positive aspects of their life, these DHH streamers believed that more people will acknowledge their efforts and talents, and this could change the public's stereotypes of the DHH community.

\subsubsection{\textbf{DHH Culture and Identity Promotion}}

Some DHH streamers had a more direct call to change people’s perceptions of the DHH community or help DHH people. They achieved this by sharing knowledge related to the DHH community. For example, P1 livestreamed interesting stories about the sign language culture, and P3 shared DHH welfare policy. Teaching sign language was a key element in P1 and P3’s livestream. P1 talked about his motivation,
 
\begin{quote}
    \emph{``I hope more people will realize that understanding sign language can benefit both DHH and non-DHh people. First, sign language can be promoted through We media to encourage more people to learn. The DHH community has communication needs because they need to be able to communicate with people without DHH in certain emergency situations. I hope the public can realize that hearing loss is not just a disability group issue. Older people also have mobility problems and hearing loss. Sign language can help the elderly with hearing loss to communicate and prevents dementia.''}
\end{quote}

P2, who livestreamed sign language instruction, laws related to people with disabilities, DHH community culture, and psychology, mentioned his motivation behind:

\begin{quote}
    \emph{``Most deaf people I have encountered feel inferior because of their hearing conditions. I hope I can help destigmatize those negative images left for the DHH community. I hope that by livestreaming DHH-related knowledge, more people with hearing loss will be aware that we too can live fulfilling lives. Livestream will allow me to help them gain self-acceptance and confidence.''}
\end{quote}

\subsection{\textbf{Livestreaming Practice (RQ2)}}\label{findings: RQ2}

In our interviews, we asked for details about the forms of interaction that participants used when livestreaming. We also asked what they did to attract and retain viewers.

\subsubsection{\textbf{Interaction Format}}
The highly diversified forms of interaction when livestreaming was one of the most notable findings.
\begin{enumerate}
    \item \textbf{Ways of Communication.}
      The ways to engage with viewers were extremely visual-demanding. Sign language was the most commonly used way of communication. 14 out of 15 participants used sign language. For example, P14 and P15 said that sign language made communication easier. P9, P14, and P15 chose sign language as the only means of communication when livestreaming. Some enjoyed performing sign language, as P9 said,\textit{`` Sign language is very rhythmic and can attract viewers. My viewers find my sign language lively and cute.''}
      
      For many streamers, sign language was used in combination with other forms of communication. One of the most important supplements was text, either text written on a writing pad (see Fig. 2, P4) or types and displayed on a screen (see Fig. 2, P5). Streamers typed or wrote down a few words or one or two simple sentences and held the pad or screen towards the camera so that their viewers could read the content on it. 
      
      An interesting and important finding was that the ways of interaction with viewers strongly determined and were determined by the viewer characteristics. P14 and P15 mentioned that when sign language was solely employed during livestreaming, almost all viewers were DHH. P4, P8, and P12 noted that text-based communication attracted more viewers, most of whom were not DHH. Therefore, when DHH streamers decide to attract more viewers who are usually not DHH, they use text communication as frequently as possible. One of our participants, P7, explained her decision of using a writing pad, \textit{"DHH people generally have lower income than people who are not DHH. I want more rewards from my viewers, so I need more viewers who are usually not DHH. Therefore, I often use a writing pad to write what I want to say. And everyone, including non-DHH viewers, can understand."} P13 preferred using a writing pad as the main means of communication for similar reasons.
      
      This advantage brought by a text-based interaction with the viewers had the cost of losing DHH viewers. Several participants mentioned this point when asked why they consistently use sign language as the only way of communication. \textit{"DHH streamers who use writing pads just want money from viewers who can hear. This way of communication is not for DHH viewers, and DHH viewers will immediately lose their interest"}, said P14, who only used sign language to interact with and retain his DHH viewers. P1, explained why he did not use a writing pad,

    \begin{quote}
        \emph{``I do not consider using a writing pad like some other streamers. Maybe because most of their viewers are not DHH, and their good looks make it easy for viewers to not to get sick of them, so their viewers can patiently read their words. I am not good looking, and I don't have many non-DHH viewers. No one will read my words on a writing pad.''}
    \end{quote}

    Unlike P7 and P13, who intentionally used a writing pad, P10 used text messages because he had no choice. He only livestreamed video games, so the only way of engagement with viewers was to type some messages in the comment box during the game break period when his hands were free. We will go back to this point when talking about the difficulties in multi-tasking.
    
    Nonetheless, most DHH streamers combined writing and sign language to interact with their viewers. Writing usually played an assistive role and was used when other people do not understand sign language. For example, P4 said he would use a writing pad when he co-performed ("\textit{lianmai}") with other streamers who are not DHH. Notably, even within the DHH group, not everyone can understand sign language, as P8 said,

    \begin{quote}
        \emph{``Most of my viewers are hard of hearing, and some of them understand sign language poorly or not at all. Although I mainly use sign language when livestreaming, some viewers may not understand what I say. They then type messages like "cannot understand" in the comment box. In this situation, I use text to facilitate communication.''}
    \end{quote}

    P12 used sticky notes instead of a writing pad because sticky notes don't need extra cleaning after writing. He also used large-sized marker pens to make his words easy to read. P11 sometimes replied to his viewers by texting in the comment area. \textit{"I usually use sign language, but I also type to reply to viewers’ messages in livestreaming chat box."}
    
    Some streamers who were able to speak or lipread use Putonghua spoken language to interact with their viewers. Some of our participants (P1, P3, P5 and P6) occasionally used spoken language to assist in communication. 

    P2, the educational streamer, believed that the more viewers, the better. He had the broadest range of communication means. Besides Putonghua spoken language and sign language, he also provided captions for livestreaming so that more people could understand and participate. The closed captions were automatically generated by the conference platform Tencent Meeting based on audio.

    \item \textbf{Co-Performance.} Co-performance was popular among DHH streamers. Eight participants mentioned that they co-performed with others. As the third and fourth pictures of Fig. \ref{fig:lianmaiVSPK} show, PK was a function provided by most livestreaming platforms. PK had two forms. One was that the streamer can PK with a specific streamer. In this case, the two streamers needed to follow each other to be friends on the platform. Alternatively, the platform could randomly assign one streamer who was online to be connected. The two streamers usually competed with each other by finishing a small task proposed by either of the streamers and their viewers could show their support when they livestreamed the task completion process. It usually took no more than three minutes for a match in PK. The one with more viewer support (virtual gifts or likes) at the end of the PK would be the winner. A bar of the realtime viewer support status, colored separately for each streamer, was displayed at the top of the livestreaming window, as shown in the last two pictures of Fig. \ref{fig:lianmaiVSPK}. \textit{Lianmai} was a more flexible form of co-performance. \textit{Lianmai} allowed multiple streamers to be connected. It had no time and participants limitations. As said, many of our participants co-performed. %P9 loved to perform sign language comic dialogue with her grandpa. \textit{"My viewers like my performance with my grandpa. They all say he is lovely!"}, said P9. Most DHH streamers chose to co-perform with familiar streamers. 
    
    P1, P3, and P4 chose familiar streamers who were online when they livestreamed. Others, such as P7 and P13, randomly co-performed (mostly random PK) with streamers who were recommended by the livestream platform.
\end{enumerate}

\subsubsection{\textbf{Practices to Maintain Popularity}}
 DHH streamers utilized various measures to retain and attract viewers. 
\begin{enumerate}
    \item \textbf{General Strategies.} Some strategies were common and well-known among streamers. Increasing the number of viewers was important. For instance, night hours were preferred because more people are available. Streamers also liked to co-perform with famous streamers who had many subscribers so that the less famous ones could attract many new viewers. Some DHH streamers also used advertising posts on various social media platforms to promote their livestream account. It was also important to engage with existing viewers as much as possible. DHH streamers responded to as many messages as they can or take suggestions from their viewers to keep the viewers engaged. 
    
    \item \textbf{Additional Labour.} Some interactions between DHH streamers and their viewers were distinct. DHH streamers and their viewers needed to make some extra effort to interact with each other. Some employed additional apps or devices. For example, P2, who promoted knowledge and culture related to the DHH groups and wanted to attract as many people as possible, used two livestream platforms on two devices when livestreaming. He explained,

    \begin{quote}
        \emph{``I have researched for a long time and finally found a hybrid pattern. Bilibili is used for screen sharing and allows viewers to quickly join the livestream by clicking on a hyper link directly without downloading any specific software. On the other hand, Tencent Meeting can automatically generate captions. By combining these services provided by the two platforms, our livestream provides three modes for information output: Putonghua spoken language, sign language, and closed captions.''}
    \end{quote}
    
    Some streamers changed their ways of communication according to their viewers' comments. As illustrated by P6,
    
    \begin{quote}
        \emph{``I can identify DHH viewers and viewers who are not DHH. I know most of my subscribers and am quite familiar with their IDs. I know their hearing conditions. But when I see new viewers, whose IDs are not familiar, I have to focus on what they say in the comment box. I usually use sign language when livestreaming. So it happens that new viewers enter comments like "encrypted video calls?", and I immediately realize that this person does not understand sign language. I will then talk a bit so that everyone can understand.''}
    \end{quote}
    
    Viewers can inform the DHH streamers about the livestreaming content. For example, the video game streamer P10, played background music when livestreaming, and he said,
    
    \begin{quote}
        \emph{``I am worried that my viewers will get bored if I livestream in silence, so I add background music. But I can't hear and don't know what songs are good. So, I randomly choose a song from an online list of the most popular songs and wait for my viewers to tell me whether it is melodious. They will ask me to switch songs if I happened to pick an unpleasant song.''}
    \end{quote}
    
    Viewers could also facilitate the livestreaming process. P13, who had many non-DHH viewers, always got help from them when she co-performed with non-DHH streamers. She said \textit{"My non-DHH viewers sometimes translate spoken language of the other streamer into words in the comment area. This helps a lot during co-performance."}

    \item \textbf{Identity Disclosure.} We noted that all of our participants somehow disclosed their DHH identity through user IDs, public profiles, or short videos. Five of them had DHH-related words, such as "dead voice," "deaf," "muted," “silence,” or "hard of hearing", in their IDs. Seven participants provided detailed descriptions of their parthenogenesis and hearing conditions in their public profiles. All participants used sign language or name their published curated videos with titles such as "deaf girl" to reveal their DHH identity.

    Streamers disclosed their DHH condition for various reasons. P14 said \textit{"I hope to attract DHH friends and find the community by including the word "deaf" in my ID"}. Several DHH streamers (P7, P11, and P13) mentioned that the DHH identity could be leveraged to attract more viewers because many people were curious about them. P8 said she planned to ask a non-DHH friend to act as her lover when livestreaming. She explained, \textit{"Although it's a publicity stunt, the romance between a DHH person and someone who is not DHH is quite intrigued and may bring more viewers."}
    
    \item \textbf{Entertaining Sign Language Performance.} DHH streamers also found different ways to make their sign language communication interesting to engage their viewers. Entertaining sign language performance was an important element of livestreaming. For example, P11 performed songs in sign language by asking his viewers to choose songs. He said, \textit{"…they (the viewers) will engage and participate in my livestream. Normally I will get more rewards if I do this."} P8 sold products on livestreaming platforms. She said, \textit{ "To demonstrate the functions of products, I make exaggerated facial expressions and hand gestures. My viewers find this amusing."} She added, \textit{"Many of my fans like to watch me mimicking foreign expressions. I like the American ones the most, they look exaggerated. Facial expressions are used in sign language. Every nation is unique."} P9 performed comical dialogues in sign language with her grandfather. \textit{"My viewers like my performance with my grandpa. They all say he is lovely!"}, said P9.
    
    \item \textbf{Other Online and Offline Activities.} 14 out of 15 participants posted curated videos on the livestream platform. P1, P9, and P12 streamers gave viewers access to their private social media accounts, such as their WeChat accounts so that the streamers and viewers can communicate even when their livestream was not on. This intensified communication between streamers and viewers might also increase viewership and retain current viewers. Some streamers held offline events. As commented by P2,
    
    \begin{quote}
        \emph{``Events that take place offline can also attract viewers. We have a public WeChat account where we post offline events information such as sign language performance in subway stations and sign language booths on the street. We have also held picture book events and offline sign language classes. Another strategy to increase the number of new subscribers is to post videos.''}
    \end{quote}
    
    As an employee of the sign language culture company, P4 received support from his employer in organizing offline events. In July 2022, he participated in a sign language competition and won the top award. After becoming the subject of several news stories, the number of his subscribers increased rapidly.

\end{enumerate}

In summary, we identify some common and distinct practices of DHH streamers, especially sign language users, when they interacted with their viewers and co-performers.

\subsection{Challenges and Some Mitigation Strategies (RQ3)}\label{findings: RQ3}
We identify four types of challenges with the use of the livestream platforms. They were categorized as challenges related to technology, interaction, content moderation, and social stereotypes. 

\subsubsection{\textbf{Technical Challenges}} 
Technical challenges was one of the most cited issue.
%Although Captions qualities\cite{Li2022} have been problems to DHH people for a long time in online videos, most video platforms have Automatic Speech Recognition (ASR) technology to generate caption\cite{AutoC}.
\begin{enumerate}
    \item \textbf{No Captions.} Most livestreaming platforms were designed for people who can hear and speak. Livestreaming platforms we knew do not offer real-time audio-to-text captions. The lack of captions made communication extremely difficult when DHH streamers co-performed with non-DHH streamers.

    Some streamers used the captioning services provided by smartphones. The AI subtitle function on HUAWEI devices was popular among our participants. When a HUAWEI smartphone was placed near the audio source and the function button was pressed, captions would be displayed on the screen. P7 used a second HUAWEI smartphone to collect captions when she was co-performing with other non-DHH streamers. But, the quality of this real-time captioning tool needed to be improved. P4 said \textit{ "The smartphones consistently provided inaccurate translations."}
    
    As noted, most of the real-time captioning services were based on speech recognition. However, most DHH streamers did not speak and rely heavily on sign language. Therefore, they could not use this service to translate their own livestream into text for those who did not understand sign language. This created a substantial communication gap between DHH streamers and non-DHH people, leaving negative experiences in their interactions with others on those platforms. P14 told us that he did not communicate with people who are not DHH when livestreaming because \textit{``They are so difficult to communicate.''} P9 tried once to co-perform with a non-DHH streamer, and she described her feelings as: \textit{``I will never accept any co-performing invitations anymore. I feel that co-performing with non-DHH streamers is very exhausting.’’} P11 added that it was challenging for him to communicate with streamers who are not DHH, thus he never co-performed with any of them. The result that DHH streamers are trapped within their DHH community is a problem for some. For example, P4 mentioned that it was difficult for him to find other streamers who perform sign language hip-hop because \textit{"The DHH community is too small to find people with similar interests."} This issue also substantially limited the number of their viewers. 
    
    DHH streamers urgently needed a service that can translate sign language into texts in real time. As P6 said,
    
    \textit{ "The livestreaming platform would benefit from AI-generated subtitles that could translate sign language. Then I wouldn't have to communicate with sign language and speech at the same time."} 
    
    P3 further added that the translation of sign language into captions was not only important for people who were not DHH to understand but also critical for those who were DHH but did not understand the sign language used by the DHH streamer,
    
    \begin{quote}
        \emph{`The caption is crucial. Some individuals are unable to use sign language. There are some communication-related problems. Case 1: unable to speak but able to use sign language; Case 2: able to speak but unable to use sign language; Case 3: unable to use sign language and unable to speak. Because using sign language and  writing on a tablet at the same time is impractical, having sign language-to-text captions is important to communicate in Case 2 and Case 3.''}
    \end{quote}
    
    Although there were some ongoing studies about the Chinese sign language recognition \cite{LSign}, P1 stated that services provided in the current App market had low accuracy and were difficult to use.\textit{ "It was impossible to achieve a high accuracy level for sign language - caption translation. There were so many local expressions in the Chinese sign language! This technology has a long way to go."}
    
    \item \textbf{Small or Low-quality Images.} The use of sign language relied heavily on the size and quality of livestreaming images. The small window size and packet loss all caused problems for DHH streamers to interact with others.
    
    The small size of the livestreaming window was a critical issue. In those livestream platforms that the participants used, the window size was fixed. According to P1, P2, and P3, when multiple streamers joined in the same window to co-perform, each of the screen of a single streamer got smaller and smaller. Then, there was not enough space to read sign language. 
    
    As introduced, the livestreaming co-performance had two forms. One was PK and the other is \textit{lianmai} (please refer to Fig. 1 with the first two pictures to be the lianmai windows and the last two to be PK windows). As said, \textit{lianmai} was less restrictive than PK because there was no time limit or participant restrictions. However, DHH streamers tried their best to PK even if they had only three minutes and needed to pay extra efforts to add other streamers as friends before PK. The much smaller size of the livestreaming window for one of the streamers in \textit{lianmai} was the key reason, as illustrated by P1,
    
    \begin{quote}
        \emph{`Sign language can still be clearly seen and understood when streamers participate in PK when they share the window size equally. When \textit{lianmai}, the guest streamers' windows will shrink, and we cannot see their sign language clearly. In PK, you can clearly read the sign language because the window is of the same size. PK with a specific person needs each other to be friends. If someone asks me to lianmai, I will ask whether he or she is willing to subscribe me. If it's okay, I'll subscribe him or her as well, and then we can become friends. We can PK.''}
    \end{quote}
    
    Packet loss happened frequently in livestream, making communication in sign language difficult. For streamers who were not DHH, viewers could still have access to their speech to comprehend the content. However, for DHH livestream viewers, no information could be relayed if the livestreaming frame was lost or unclear. Packet loss seemed to occur when the Internet speed was low, as mentioned by P2. P6 also talked about the delay effect due to the slow transmission of images over the internet, 
    
   \begin{quote}   
        \textit{ “When I am done with one topic and start with a new one, my viewers only start laughing about the previous topic (They would type "Hhh" in the comment area). That’s too late!”}
     \end{quote}
    
\end{enumerate}

\subsubsection{\textbf{Challenges Related to Interaction}}
DHH streamers encountered numerous issues interacting with their audiences and other streamers. Their specific interactional difficulties were described in the following four subsections. 

\begin{enumerate}
    \item \textbf{Difficulties in Multi-tasking.} For many DHH streamers, sign language was the most effective way to express themselves. Their hands were busy and cannot be used for other activities. DHH streamers could not communicate other information when they used their hands for tasks such as playing video games, demonstrating products, making artwork, etc. This difficulty in multi-tasking made it less effective for DHH streamers to timely interact with their viewers. P10, the video game streamer, mentioned, 

    \begin{quote}
        \emph{``I am unable to play games and type texts or use sign language simultaneously. Game playing will be affected. Only when my roles have died in the game will I type to explain my gaming skills to my viewers.''}
    \end{quote}
    
    The delay in response to viewer comments often led to loss of viewers. As P4, who was also a video game streamer, said, \textit{``By the time I come back to read the comments, the viewers have already left.''} P8 also felt that it was difficult to respond to their viewers timely when she was selling food using sign language during the livestreaming:
    
    \begin{quote}
        \emph{``For example, when I was selling a fondant cake, I first displayed the box to my viewers before introducing the fondant's brand and price. Then I'll encourage them to click the link to place an order. I'll open the packet and the fondant so that everyone can see the filling. I'll continue to tell the taste by sign language after I've eaten it. It is impossible to simultaneously display products and use sign language for me.''}
    \end{quote}
    
    The difficulty of multitasking even affected the livestreaming content. P7, explained why she did not livestream her painting process, \textit{``When I am painting, I can't use sign language. Also, I can't speak. Therefore, I can’t communicate with my viewers (when I am painting), so they get bored and leave.''}. 
    
    \item \textbf{Sign Language Diversity.} %Natural sign language was the predominant mode of everyday communication among Chinese DHH groups.  This sign language was different from the Chinese sign language (CSL). CSL was based on the Chinese speaking language and follows Putonghua grammar. Chinese natural sign language was derived from the daily life and social context of DHH people and was therefore regional customary.
    
    The regional-specific feature of the natural sign language in China was noted in many interviews. For example, P9 used to live in Guangxi province in South China. So, she couldn't understand the sign language used in North China. P2 added that viewers from different regions of China occasionally had difficulty understanding her sign language. A high level of sign language skills was important to attract DHH viewers from various regions. As P1 said, 

    \begin{quote}
        \emph{``Different regions of China have their own sign languages. I have been using sign language for a long time and am interested in learning sign languages of different regions. Therefore, I use sign language in many different ways. For instance, there are three or four different sign languages to spell a name, so I will show them all until my viewers understand what I am saying.''}
    \end{quote}
    
    P3, a sign language teacher, said \textit{``There are different sign languages in different regions, and there are also sign languages in other countries. In general, I use a variety of sign languages when livestreaming.''}
    
    \item \textbf{Fixed Viewer Group.} As discussed above, the way of communication could affect who the viewers are and what they livestream. Once a DHH streamer’s viewer community was established, many DHH streamers would stick with their current communication method to retain those viewers.  P1 explained why he kept using sign language when livestreaming,
     \begin{quote}
         \textit{``I want to attract non-DHH groups. But making a change was challenging.''} 
      \end{quote}
      
      P15 further elaborated on this point,

    \begin{quote}
        \emph{``You have to write on a tablet or a pad if you want more non-DHH viewers. However, there will be fewer DHH viewers if I utilize a writing pad. A DHH streamer using a writing pad and pen will not attract DHH viewers. They will go away. It is very likely that you will lose many long-time DHH viewers and fail to attract new viewers who are not DHH. The DHH audience community was initially set up in this manner. Therefore, it is very difficult to change it later.''}
    \end{quote}

\end{enumerate}

\subsubsection{\textbf{Challenges Related to Content Moderation}} Participants talked about their unpleasant experiences related to a misunderstanding about their livestreaming content.

\begin{enumerate}
    \item \textbf{Internet traffic-limiting.} Two participants (P7 and P9) suspected that their livestreaming account was suspended. Although they did not receive a formal notification of account suspension, no new viewers had visited them from the Douyin home page for a long time. P7 said, 

    \begin{quote}
        \emph{``My Douyin account was suspended for a very long time. I asked every DHH streamer I know and their accounts were all restricted. Their in-time viewership was down, and in fact all their viewers were regular viewers. There were no new viewers. Maybe we were reported by the platform housekeeper or by the anti-fans, or maybe the livestream platform stopped bringing new online traffic to the disabled group because more people are pretending to be deaf recently.''}
    \end{quote}
    
    \item \textbf{Lack of Sign Language Moderators.} Currently, content moderators on livestreaming platforms did not take into account of communication in sign language. P1 told us that due to the lack of sign language content moderators, some offensive and profane content was seen in some DHH livestreaming.
    
    Some DHH livestreaming rooms were banned by mistake. P3 recalled that she had once showed her middle finger when livestreaming and her livestream room was immediately banned, even though she actually just meant to say the word “middle finger”, as she said, \textit{``I didn’t mean to be rude!''}
    
\end{enumerate}

\subsubsection{\textbf{Challenges Related to Social Distrust and Discrimination}} 
Twelve of our participants told us that their unpleasant livestreaming experiences were related to their DHH conditions. The distrust of their DHH condition and discriminative language and behavior towards people who were DHH were the two main sources of unpleasant experiences. P1 shared his experience, 

\begin{quote}
    \emph{``I tried to co-perform with streamers who are not DHH, but they all cursed me. They said I was just faking my DHH state to get sympathy. They said that I was repulsive and lying.''}
\end{quote}

P10 also mentioned that some viewers did not believe that he was DHH. He said \textit{"Some viewers said it was a pity I couldn't talk or hear in the comment area. Or ‘do you pretend?’ ‘Unable to hear?’ ‘Why are you silent?’ Many people claim that I pretend to be DHH and that I am shameless."}

P4, the hip-hop sign language performer, said that whenever he tried to co-perform with streamers who were not DHH, he was always turned down. He said, \textit{"They said I was a liar and showed me a middle finger."} 

There seemed to be a strong stereotype towards DHH people, labeling them as slow and incapable. P7 added that some streamers shut the co-performance window once they saw her writing pad. \textit{"Once they see a writing pad, they know that I am unable to hear and speak. Most streamers don't want to co-perform with DHH streamers."}

P9, an owner of a nail salon, said \textit{"I was delighted to share the opening of my nail salon during my livestreaming. My viewers thought it was terrific that DHH people could open a store. However, many of them doubted me, so I showed them my business license."} 

P2 also shared his struggle in promoting DHH identity and culture. He said \textit{"Doing educational livestream about the DHH identity is difficult. Some of my viewers said negative things, such as, ‘What is the use of doing this?’ ‘Who will watch it?’ ‘Deaf is deaf!’ ‘It is useless (for DHH people) to go to college!’"}

P11 talked about the general impression towards DHH people based on his livestreaming experiences, 

\begin{quote}
    \emph{``Many people doubted me. I guessed they had stereotypes about DHH people. Deaf = poor, stupid, and illiterate!''}
\end{quote}

Those negative comments made some streamers depressed. P2 said\textit{"It requires a lot of courage. This kind of brutality worries me. I don't like receiving criticism. It was quite depressing when hearing about those (unpleasant) experiences of other DHH streamers."} P5, a WeChat group streamer whose most viewers were her friends, was concerned about having viewers that she does not know, \textit{"Because I think I need to meet people who are not DHH, and that will put extra pressure on me."}

DHH streamers had different responses towards those doubts and discriminations. Some would clarify. \textit{“When I see these doubts, I usually post clarifications in text. Sometimes, my viewers explain for me.”} said P11. P12 also said he would reply in text when he saw questions such as “Why don’t you speak?” Some (P6 and P10) chose to ignore those comments because it was exhausting to respond.

On the livestreaming platforms, some DHH streamers had their disability certificates displayed on their public profiles. We asked our participants whether they would do this to avoid criticism. None of them was willing to do so. P1 said,

\begin{quote}
    \emph{``They said I was repulsive and lying, possibly because I don’t have a disability certificate on my livestreaming profile, even though many DHH streamers have it displayed. … It (the certificate) is not something I want to highlight. It is all about looking down on people with a disability certificate. I feel uncomfortable to be made to be discriminated.''}
\end{quote}

P3 expressed similar feelings, 
\begin{quote}
    \emph{``Some people present a disability certificate, and they can get some money by getting others’ sympathy. I am ashamed of this behavior. Why do we need to show our disability certificate? ... We don’t want to be pitied. I don’t see this as a way to gain sympathy by presenting a disability certificate.'' }
\end{quote}

A minority of the DHH streamers (P14 and P15) said that they did not encounter any identity-related issues when they livestream. Interestingly, P14 and P15 only used sign language during the livestream. Probably because of all their viewers were DHH due to this single way of communication, the streamers' DHH identity was not something in question.

\section{DISCUSSION}

In this paper, we interviewed 15 DHH streamers who livestream a variety of content on different livestreaming platforms. We focused on their motivations for livestreaming, practices on livestreaming platforms and reasons behind, and challenges with the current livestreaming interface. Our findings highlight the distinct interaction model of DHH streamers on livestreaming platforms and fill an important gap in the literature on accessibility. Overall, DHH people employed multiple ways to interact with viewers and co-performers, and this way of interaction was embedded in their desire to change stereotypes towards them and to overcome barriers of communication through natural sign language, featured in their distinct sign language culture, and was shaped by the design of the livestreaming platforms that strongly demand audio inputs. This was pivotal for us to understand how they interact on livestreaming platforms and to give some useful design suggestions.

\subsection{Design to Facilitate DHH Streamers' Communication with Audience}
The great communication barrier between people who understand sign language and those who do not characterizes DHH streamers’ choice of communication and viewer pool. DHH streamers need to trade off between using sign language and texts (e.g., a writing pad) depending on whether they want to engage DHH viewers or other viewers. Many participants mentioned that they chose not to use a writing pad because they did not want to lose their DHH viewers. Similarly, those who want to attract people who are not DHH minimize the use of sign language at the expense of losing DHH viewers. The current livestreaming platform does not offer DHH streamers the opportunity to attract both viewer groups. To solve this problem, we identify some areas for improvement and offer some suggestions below.

First, successful communication with sign language and texts depends heavily on the size of the livestreaming window. As mentioned in Section 4.3.1, DHH streamers prefer \textit{PK} over \textit{lianmai} because the window size in \textit{lianmai} is small and cannot be adjusted.   
We thus suggest that the livestreaming window should be adjustable. The communication experience can also benefit from the customization of the speaker detection feature, which automatically expands the speaker's window when triggered by an audio effect \cite{ene1,disbu2022,4064524}. Specifically for DHH streamers, we propose the introduction of a dedicated gesture recognition feature, as previous work has developed several ways such as used pixels \cite{Angel2022}, optical flow features \cite{Moryossef2022} and video features of sign language \cite{Shipman2017} to detect signing that can spotlight the speaker during livestream.

Sign language also places high demands on video quality, which is strongly influenced by network bandwidth. Especially during co-performance, loss of frames hinders timely communication for DHH groups more than for non-DHH streamers. In scenarios with poor network connection, the lost or blurred frames of video can severely impact the sign language-based communication, while hearing people could still rely on audio communication, which has lower demands on the network quality. The importance of network bandwidth was also mentioned in the study on the use of video calls by DHH users \cite{Tang2021}.

%the DHH broadcasters' livestreaming quality is significantly impacted by the order of their sign language actions. Their experiences will be strongly influenced by video quality of audience window (\textit{lianmai}’s window). To ensure DHH groups could have better user experiences, we advise livestreaming apps should have some solutions to deal with video quality during \textit{lianmai}. If the video images are lost due to limited bandwidth, no information can be transmitted for streamers using sign language only, while hearing people can still rely on audio messages for communication. %

Text communication is an important medium to bridge the communication gap between people who use and who do not use sign languages, but also brings the community of DHH people who use different natural sign languages in different areas of China. Many of our participants requested real-time subtitles for livestreaming in sign language. Although there are some developments in sign language recognition or even translation with the advances in machine learning techniques \cite{LSign,AccSignL,wu2015real,3DSignL}, sign language recognition in China remains to be a serious challenge because the use of region-specific natural sign languages is widespread among Chinese DHH groups and there are few corresponding datasets for training effective sign language recognition models. 

The lack of the inclusive design also makes the current livestreaming platforms unable to moderate the livestreaming content in sign language. This leads to profane or vulgar content in sign language or wrongly suspended user accounts. In addition to the calling for sign language recognition techniques, we also suggest that content moderators have more deaf awareness or receive some sign language training in their work. The regional-specific feature of the local natural sign languages makes this issue even more challenging in China.

\subsection{Design for Inclusive Identity Management}
Breaking the stereotypes towards DHH people is an important driver for our DHH streamers to livestream and to continue livestreaming. Among our participants, they seldom put a great effort to hide their DHH identity, and all of them disclosed their identity on the platform in some ways (but they did not like publishing their disability certificate). This observation is in stark contrast to findings of the previous study on BLV streamers, where the BLV streamers actively downplay their disabled identity online to gain more viewers who are not BLV \cite{Rong2022}. The unique mode of communication among DHH people, in which sign language plays a central role, may be the reason why the DHH livestreaming community remains to be relatively closed or even self-sufficient.

Nonetheless, the distinct mode of communication on livestreaming also makes DHH streamers a target for trolls and discriminatory language. The same was observed for BLV streamers \cite{Rong2022}. In particular, doubts and misperceptions often surface in the comment box or when DHH streamers co-perform with unfamiliar streamers. Some of our participants even reported experiencing traffic limiting. They suspected that viewers out of their DHH community regarded them to be fake DHH people and report them to the platform. DHH streamers face the dilemma of minimizing such mistrust by publishing their disability certificate and not advocating that status with the intention of not being pitied. 

Livestreaming platforms should develop a verification process to certify people who are disabled if they apply. Currently, this verification service is available only for public figures, government officials, and other individuals with a certain level of visibility on Kuaishou and several livestreaming platforms. For people with disabilities, this service should be provided given the significant suspicions and misconceptions about them. 

\subsection{Traffic Limiting Issue}  
Our participants reported the experience of having no new viewers or subscribers to visit them for a long period of time. Different from the findings of the earlier study about BLV streamers where the interviewees attributed this to algorithmic bias \cite{Rong2022}, our participants suspected that their accounts were reported by viewers with strong discrimination and distrust toward DHH groups. Although we cannot verify what are the true reasons for this experience, we would like to highlight again that traffic-limiting is a common problem for minority groups. This resonates with prior work which identified the underlying traffic limiting for LGBT streamers \cite{Cai19,Uttarapong2021} and people with disabilities \cite{linessocial}, including facial disfigurement, Down syndrome \cite{robertson2019tiktok}, and eye disorders \cite{biddle2020invisible} in TikTok.

Current algorithms that value audio input and the lack of consideration of the characteristics of DHH streamers during the account verification process may further marginalize their position on livestreaming platforms. Further work is needed to verify the presence and extent of algorithm bias against this group.

\subsection{Design with DHH Group's Early and Active Involvement}
Current livestreaming apps are not designed from the ground up to meet the needs of people with disabilities. An inclusive and more accessible design should take into account the accessibility studies perspective, which gives disabled people agency and control from the beginning of a technology development phase \cite{hofmann2020living}. A new livestreaming platform developed for DHH and hearing groups, with DHH individuals involved at every stage of development, has the potential to provide DHH streamers and their viewers with new and improved user experiences. From the start, DHH users and their needs and goals should be included. For example, a wide-ranging and open-ended view of the potential solutions from sign language livestream communication should be included \cite{lazar2017research}. Regardless of the feasibility of the idea with existing livestreaming programs, DHH users can suggest a perfect system that most appeals to them \cite{lazar2017research}. Finally, the comments and subjective reactions of DHH users are important as the livestreaming program improves and evolves \cite{lazar2017research}. It is essential to conduct usability tests and empirical studies with the inclusion of DHH users. We look forward to a future livestream app that includes DHH users in its development from the beginning.

\section{Limitations and Future Work}

\subsection{Limited Livestreaming Platforms}

We investigated the livestreaming experiences of DHH streamers on multiple Chinese livestreaming platforms such as Kuaishou, Douyin, Bilibili, Tencent Meeting, Wechat Group Live, and Huya. We expect that our results should be generalizable to most livestreaming platforms. However, other livestreaming platforms may vary in their interface, promotion algorithms, or user account management system that may affect the user experiences. 

\subsection{A Limited Recruitment Channel}
We reiterate that all of our participants came from a single recruitment channel---online posters on Xiaohongshu, although we tried multiple channels. It is possible that the current findings reflect only a small proportion of DHH streamers’ livestreaming experiences. However, the wide range of the livestreaming content and the use of multiple livestreaming platforms, as well as the diversity of livestreaming communication methods may indicate that the issue of having a highly selective and specific sample is not very serious. 

\subsection{A Limited Coverage of the DHH groups}

Our study is primarily concerned with sign language users rather than the broader DHH population, many of whom do not use sign language. Therefore, this study focuses heavily on how sign language users use and interact with current livestream platforms. Similar to other streamers, our participants were young. Only 3 of them were older than 30. We are aware that many people are DHH at a relatively old age, fluent in Putonghua, and do not use sign language. Exploring their experiences and challenges with livestreaming is an interesting and important extension of the current study. Nonetheless, our study stands out in its focus on the livestream experiences of sign language users as well as those who can both speak and use sign language. We believe their perspective on current livestream apps makes an important contribution to the literature.

Besides, our participants' livestreaming histories range from two weeks to four years. P12 and P13, two full-time streamers, put in 6–9 hours a day of livestreaming. Most of the other participants were part-time livestreamers who did not follow a regular livestreaming schedule. Livestreaming practices and experiences should be dependent on one's livestreaming history. It would be exciting if we had longitudinal data following the same streamers over time to examine how their views and experiences of livestreaming change. In the current study, we limit our focus to the general livestreaming experiences of DHH streamers and are pleased to report that streamers with a wide range of livestream experiences all provided valuable information to our findings.

\subsection{Working with a Sign Language Interpreter}
Finally, we would like to reiterate that all four researchers are not members of the DHH community. We conducted our interviews primarily via text messaging. For the video interviews, we were generously assisted by a sign language interpreter. The researchers' lack of sign language skills can make communication with DHH streamers less efficient and accurate, resulting in the loss of some nuanced insights and even power-imbalance during the interview. Nonetheless, we did not find that the video interviews offered less detail compared to the text interviews. We also did not ask any sensitive or judgmental questions in the interview. All interviewers were initially contacted via text messages and could communicate with us via text anytime. As evidenced by the results and the beginning of the discussion sessions, we were impressed by our participants' strong desire to break stereotypes and their pride in their identities. The purpose of this study is to understand the livestream experiences of DHH groups, and no evidence showed that our results were biased due to our limited sign language skills. Nevertheless, future work should benefit from involving researchers with extensive knowledge of sign language in China. In fact, we are in the process of recruiting a student researcher who is DHH.

\section{CONCLUSION}
\label{CONCLUSION}
This qualitative study depicted a rich and insightful picture of DHH streamers’ motivations, practices, and challenges on multiple livestreaming platforms. As one of the first works on the livestreaming experiences of DHH people, our study highlighted the importance of DHH identity in shaping their livestreaming motivations, choices of commutation methods, ways of interactions with viewers and other streamers, and challenges they had on the livestreaming platforms. Many of our participants livestreamed with the hope of changing people’s perception or stereotype towards the DHH group. %The finding that livestreaming platforms offer new opportunities for DHH people to advocate their identity and culture is also noted a few earlier work about how streamers with disabilities feel empowered to inspire the others and change the stereotype towards this group \cite{anderson2021gamer, johnson2019inclusion}. 
The two main forms of communication during livestreaming were sign language and text, and DHH streamers either use one or a combination of them depending on their viewer characteristics. The difficulty of engaging both DHH viewers and those who were not DHH at the same time was prominent. When they livestreamed, DHH streamers often encountered doubts, misperceptions, and sometimes discriminate language in the comment box or from other streamers who were not DHH. For co-performances with multiple streamers, the small window size made sign language communication difficult. Based on these findings, we offered some suggestions to improve the accessibility of livestreaming platforms in the areas of identity management, user interface/interaction windows, sign language recognition and translation. This study sheds lights into how DHH people embrace the new opportunities brought by livestream. During this process, they had developed their distinct ways to interact with viewers and other streamer.%s, while there remained to be many challenges and problems. We hope that this work can inspire more studies to understand the digital experiences and demand of marginalized groups, and that our suggestions for livestream platform design can benefit the DHH livestreaming community. 

%This qualitative study of DHH livestreamers painted a general picture of DHH people using livestreaming platforms.Through our semi-structured interview with 15 DHH livestreamers, we pinpointed their motivations included making money,building social connection,presenting themselves and their performances as well as education.A deeper purpose for the latter two motivations were to de-stigmatize DHH community.More importantly ,we discovered how DHH livestreamer communicated on nowadays' livestreaming platforms which were designed for people who can see and hear.Sign language,written language and lip language were used to achieve accessibility.To attract more viewers and create more performance ornamental value,strategies such as various kind of sign language performance were used.As a minority group,DHH livestreamers encountered technical,social,interactional and moderational challenges.We presented some design suggestions of platforms in order to make DHH livestreamers more comfortable. In sum,this is a general work to reveal DHH livestreamers' motivations,practices and challenges.

\section*{ACKNOWLEDGMENTS}
\label{ACKNOWLEDGMENTS}
We thank Miss XU Mengjiao for being our sign language interpreter.

\bibliographystyle{ACM-Reference-Format}
%\balance
\bibliography{main}

%\appendix
%\input{Source/06-Appendix}

\end{document}